\documentclass[aps,preprint]{revtex4}%
\usepackage{amsfonts}
\usepackage{amsmath}
\usepackage{amssymb}
\usepackage{graphicx}%
\begin{document}
\title{Influence of field mass and acceleration on entanglement generation}
\author{Yongjie Pan$^{a}$}
\author{Jiatong Yan$^{b}$}
\author{Sansheng Yang$^{c}$}
\author{Baocheng Zhang$^{a}$}
\email{zhangbaocheng@cug.edu.cn}
\affiliation{$^{a}$School of Mathematics and Physics, China University of Geosciences,
Wuhan 430074, China}
\affiliation{$^{b}$Physics Department, Brown University,Providence, Rhode Island 02912, USA}
\affiliation{$^{c}$Wuhan Maritime Communication Research Institute, Wuhan 430205, China}

\begin{abstract}
We explore the entanglement dynamics of two detectors undergoing uniform
acceleration and circular motion within a massive scalar field, while also
investigating the influence of the anti-Unruh effect on entanglement
harvesting. Contrary to the conventional understanding of the weak anti-Unruh
effect, where entanglement typically increases, we observe that the maximum
entanglement between detectors does not exhibit a strict monotonic dependence
on detector acceleration. Particularly at low accelerations, fluctuations in
the entanglement maxima show a strong correlation with fluctuations in
detector transition rates. We also find that the maximum entanglement of
detectors tends to increase with smaller field masses. Novelly, our findings
indicate the absence of a strong anti-Unruh effect in (3+1)-dimensional
massive scalar fields. Instead, thermal effects arising from acceleration
contribute to a decrease in the detector entanglement maximum.

\end{abstract}
\maketitle

\section{Introduction}

The Unruh effect posits that a uniformly accelerated detector with
acceleration $a$ perceives the Minkowski vacuum as a thermal state at the
Unruh temperature $T_{lin}=\frac{a}{2\pi}$ \cite{unruh1976notes}. This
temperature can also be accurately determined from the transition probability
and the detailed equilibrium temperature when an accelerating Unruh-DeWitt
detector is coupled with a massless field
\cite{Hawking1979GeneralRA,takagi1986vacuum,crispino2008unruh}. Recent studies
have revealed that under special conditions, when the detector is coupled with
a massive field, the transition probability and detailed equilibrium
temperature are not directly proportional to the detector's acceleration. This
phenomenon has been termed the anti-Unruh effect
\cite{brenna2016anti,garay2016thermalization}. Specifically, a decrease in the
detector's transition probability alongside an increase in acceleration is
known as the weak anti-Unruh effect, while the associated decrease in the
detailed equilibrium temperature is termed the strong anti-Unruh effect.

Local coupling between detectors and fields is of significant importance in
various applications. When considering the first-order perturbative expansion
of the time evolution operator for such local coupling processes, the
entanglement of an initially entangled state undergoing accelerated motion
tends to decrease due to thermal effects caused by acceleration
\cite{fuentes2005alice,alsing2006entanglement,martin2009fermionic,bruschi2012particle}%
. However, in the presence of the anti-Unruh effect, the entanglement of both
two-body entangled states \cite{li2018would} and multi-body entangled states
\cite{pan2020influence} can actually increase. Furthermore, when moving to the
second-order perturbative expansion of the time evolution operator, a pair of
detectors that were initially unrelated can become entangled through local
coupling with a vacuum field, even if they are spatially distant. This
phenomenon is aptly termed \textquotedblleft entanglement
harvesting\textquotedblright%
\ \cite{reznik2003entanglement,Zhang2020circ,Gallock2021black,chowdhury2022fate,Pozas2015gta,Tjoa2020Vacuum}%
. This article specifically focuses on the entanglement harvesting process
involving detectors.

The Unruh effect and the anti-Unruh effect exhibit opposite impacts on quantum
entanglement, and their persistence in entanglement harvesting reflects
directly on detector entanglement dynamics. Recent studies emphasize the
significance of field mass in relation to the anti-Unruh effect
\cite{garay2016thermalization,wu2023conditions,yan2023reveal}. Consequently,
there is considerable interest in exploring entanglement harvesting processes
within massive fields, including considerations such as environmental
interaction effects \cite{Zhou2021nyv,chen2022entanglement}, detector energy
level gaps \cite{Maeso2022Entanglement,Hu2022nxc}, and high-dimensional
spacetime \cite{yan2022effect}. Our investigation delves into the role of
field mass in the entanglement harvesting process involving detectors within a
massive field, aiming to ascertain whether the anti-Unruh effect contributes
to detector entanglement. In contrast to previous studies of detector
entanglement dynamics in massive fields, we discuss in detail the influence of
strong and weak anti-Unruh effects on entanglement generation in the evolution
of detector dynamics. It is found that the time-delay effect induced by the
mass of the field, as well as only the weak anti-Unruh effect has an effect on
the entanglement generation in the evolution of the detector dynamics.

Typically, the Unruh effect pertains to detectors undergoing linear
acceleration, but exploring the Unruh effect in circular motion is
particularly intriguing due to the feasibility of achieving high accelerations
required for experimental verification \cite{Bell1982qr}. Furthermore, the
temperature measured by detectors in circular motion exhibits similarities but
also differences compared to the conventional Unruh temperature
\cite{Biermann2020Unruh}. Therefore, understanding the anti-Unruh effect in
circular motion and its implications for entanglement harvesting is crucial
\cite{salton2015acceleration,liu2022does}. Recent studies have examined
entanglement dynamics in circular motion within massless \cite{Zhang2020circ},
electromagnetic fields \cite{Pozas2016rsh,Perche2022Harvesting}, and other
types of world lines \cite{Bozanic2023Correlation}. In this context, we focus
on elucidating the entanglement dynamics between detectors engaged in circular
motion within a massive field, aiming to explore the contributions of the
anti-Unruh effect to this process.

This paper is organized as follows. In the second section, we introduce the
process of coupling two detectors to a vacuum field. Next, in the third
section, we compare the entanglement dynamics of detectors when coupled to
both a massless field and a massive field. The fourth section provides a
detailed examination of the entanglement dynamics of detectors in circular
motion. Subsequently, in the fifth section, we discuss the correlation between
the anti-Unruh effect and the entanglement dynamics. Finally, we conclude the
paper with a summary in the sixth section.

\section{Master equation}

We study the entanglement dynamics of two uniformly accelerated two-level
atoms, which are weakly coupled to fluctuating massless and massive scalar
fields in vacuum. The Hamiltonian of the two detector can be expressed as
\cite{audretsch1994spontaneous}%
\begin{equation}
H_{A}=\frac{\omega}{2}\sigma_{3}^{(1)}+\frac{\omega}{2}\sigma_{3}^{(2)},
\label{HA}%
\end{equation}
where $\sigma_{i}^{(1)}=\sigma_{i}\otimes\sigma_{0},\sigma_{i}^{(2)}%
=\sigma_{0}\otimes\sigma_{i}$ are operators of detectors $1$ and $2$
respectively, with $\sigma_{i}$ $(i=1,2,3)$ being the Pauli matrices, and
$\sigma_{0}$ is the $2\times2$ unit matrix. We assume that the excitation
energy of the two detectors are the same, and labeled as $\omega$. The
interaction Hamiltonian $H_{I}$ between the detector and the vacuum scalar
field can be written as%
\begin{equation}
H_{I}=\lambda\left[  \sigma_{y}^{(1)}\Phi\left(  t,x_{1}\right)  +\sigma
_{y}^{(2)}\Phi\left(  t,x_{2}\right)  \right]  , \label{HI}%
\end{equation}
where $\lambda$ is the coupling constant which is assumed to be small.

In the Born-Markov approximation, the master equation describing the
dissipative dynamics of the two-detector subsystem can be written in the
Gorini-Kossakowski-Lindblad-Sudarshan form as
\cite{gorini1978properties,lindblad1976generators,breuer2002theory}%
\begin{equation}
\frac{\partial\rho(\tau)}{\partial\tau}=-i\left[  H_{\mathrm{eff}},\rho
(\tau)\right]  +\mathcal{D}[\rho(\tau)], \label{ME}%
\end{equation}
where
\begin{equation}
H_{\mathrm{eff}}=H_{A}-\frac{i}{2}\sum_{\alpha,\beta=1}^{2}\sum_{i,j=1}%
^{3}H_{ij}^{(\alpha\beta)}\sigma_{i}^{(\alpha)}\sigma_{j}^{(\beta)},
\label{HEFF}%
\end{equation}
with $H_{ij}^{(\alpha\beta)}$ can be obtained by the Hilbert transform
$\mathcal{K}^{(\alpha\beta)}$, i.e.,
\begin{equation}
\mathcal{K}^{(\alpha\beta)}(\lambda)=\frac{1}{\pi i}P\int_{-\infty}^{\infty
}d\omega\frac{F^{(\alpha\beta)}(\omega)}{\omega-\lambda}, \label{HK}%
\end{equation}
where $P$ represents the principal value of the integral and $F^{(\alpha
\beta)}(\omega)$ is interpreted below, and
\begin{equation}
\mathcal{D}[\rho(\tau)]=\frac{1}{2}\sum_{\alpha,\beta=1}^{2}\sum_{i,j=1}%
^{3}C_{ij}^{(\alpha\beta)}[2\sigma_{j}^{(\beta)}\rho\sigma_{i}^{(\alpha
)}-\sigma_{i}^{(\alpha)}\rho\sigma_{j}^{(\beta)}-\rho\sigma_{i}^{(\alpha
)}\sigma_{j}^{(\beta)}]. \label{D}%
\end{equation}

From the master equation (\ref{ME}), it is clear that the environment leads to
decoherence and dissipation described by the dissipator $\mathcal{D}[\rho
(\tau)]$ such that the evolution of the quantum system is nonunitary on one
hand, and it also gives rise to a modification of the unitary evolution term
which incarnates in the Hamiltonian $H_{\mathrm{eff}\text{ }}$ on the other hand.

Therefore, the master equation (\ref{ME}) can be rewritten in the following
form,
\begin{align}
\frac{\partial\rho(\tau)}{\partial\tau} &  =-i\tilde{\omega}\sum_{\alpha
=1}^{2}\left[  \sigma_{3}^{(\alpha)},\rho(\tau)\right]  +i\sum_{i,j=1}%
^{3}\Omega_{ij}^{(12)}\left[  \sigma_{i}\otimes\sigma_{j},\rho(\tau)\right]
\nonumber\\
&  +\frac{1}{2}\sum_{\alpha,\beta=1}^{2}\sum_{i,j=1}^{3}C_{ij}^{(\alpha\beta
)}[2\sigma_{j}^{(\beta)}\rho\sigma_{i}^{(\alpha)}-\sigma_{i}^{(\alpha)}%
\rho\sigma_{j}^{(\beta)}-\rho\sigma_{i}^{(\alpha)}\sigma_{j}^{(\beta
)}],\label{MEHJ}%
\end{align}
where $\tilde{\omega}$ is a redefined energy gap. The related coefficients is
given as $\Omega_{ij}^{(12)}=\frac{i\lambda^{2}}{4}\{\left[  \mathcal{K}%
^{(12)}(\omega)+\mathcal{K}^{(12)}(-\omega)\right]  \delta_{ij}-\left[
\mathcal{K}^{(12)}(\omega)+\mathcal{K}^{(12)}(-\omega)\right]  \delta
_{3i}\delta_{3j}\}$, and $C_{ij}^{(\alpha\beta)}=A^{(\alpha\beta)}\delta
_{ij}-iB^{(\alpha\beta)}\epsilon_{ijk}\delta_{3k}-A^{(\alpha\beta)}\delta
_{3i}\delta_{3j}$, where $A^{(\alpha\beta)}=\frac{\lambda^{2}}{4}\left[
F^{(\alpha\beta)}(\omega)+F^{(\alpha\beta)}(-\omega)\right]  $ and
$B^{(\alpha\beta)}=\frac{\lambda^{2}}{4}\left[  F^{(\alpha\beta)}%
(\omega)-F^{(\alpha\beta)}(-\omega)\right]  $. In the above expressions,
\begin{equation}
F^{(\alpha\beta)}(\omega)=\int_{-\infty}^{\infty}d\Delta\tau e^{i\omega
\Delta\tau}\left\langle \Phi\left(  \tau,x_{\alpha}\right)  \Phi\left(
\tau^{\prime},x_{\beta}\right)  \right\rangle \label{GF}%
\end{equation}
is the Fourier transform of the scalar field correlation function
$\left\langle \Phi\left(  \tau,x_{\alpha}\right)  \Phi\left(  \tau^{\prime
},x_{\beta}\right)  \right\rangle $ \cite{Bire82}.

\section{Entanglement change}

Consider the two detectors separated by $L$ undergoing linear acceleration in
the (3+1)-dimensional vacuum field. Their worldlines are
\begin{align}
t_{1}  &  =\frac{1}{a}sinh(a\tau),x_{1}=\frac{1}{a}cosh(a\tau),y_{1}%
=0,z_{1}=0,\nonumber\\
t_{2}  &  =\frac{1}{a}sinh(a\tau),x_{2}=\frac{1}{a}cosh(a\tau),y_{2}%
=0,z_{2}=L. \label{WLL}%
\end{align}
To accurately show the evolution of the density matrix of the detector, we
need to calculate the correlation function of the massive and massless scalar fields.

For a massless scalar field,
\begin{equation}
G_{ml}^{(\alpha\beta)}(\Delta\tau)=-\frac{1}{4\pi^{2}}\frac{1}{\left(
t-t^{\prime}-i\epsilon\right)  ^{2}-\left(  \mathbf{x}-\mathbf{x}^{\prime
}\right)  ^{2}}, \label{Wightman}%
\end{equation}
where, $\mathbf{x}$ represents the spacial component, and the superscripts
$\alpha$ and $\beta$ are omitted for the simplicity of the formula. The
Fourier transformation of the correlation functions are
\begin{align}
F^{(11)}(\omega)  &  =F^{(22)}(\omega)=\frac{1}{2\pi}\frac{\omega}%
{1-e^{-\frac{2\pi\omega}{a}}},\nonumber\\
F^{(12)}(\omega)  &  =F^{(21)}(\omega)=\frac{1}{2\pi}\frac{\omega}%
{1-e^{-\frac{2\pi\omega}{a}}}\frac{\sin\left(  \frac{2\omega}{a}\sinh
^{-1}\frac{aL}{2}\right)  }{\omega L\sqrt{1+a^{2}L^{2}/4}}.
\end{align}

For the massive scalar field,
\begin{equation}
G_{ms}^{(\alpha\varrho)}(\Delta\tau)=\int_{m}^{\infty}d\omega_{k}\frac
{\sin\left(  \sqrt{\omega_{k}^{2}-m^{2}}\left\vert \Delta x_{\alpha\varrho
}\right\vert \right)  }{\left\vert \Delta x_{\alpha\varrho}\right\vert
}e^{-i\omega_{k}\Delta t_{\alpha\varrho}}, \label{GM}%
\end{equation}
where $\left\vert \Delta x_{\alpha\varrho}\right\vert =\sqrt{\left\vert
x_{\alpha}-x_{\varrho}^{\prime}\right\vert },\Delta t_{\alpha\varrho
}=t_{\alpha}-t_{\varrho}^{\prime}$. According to the word line (\ref{WLL}), it
is obtained \cite{Zhou2021nyv},%
\begin{equation}
G^{(11)}(\Delta\tau)=G^{(22)}(\Delta\tau)=\frac{m^{2}}{4\pi^{2}}\int
_{1}^{\infty}dx\left(  x^{2}-1\right)  ^{\frac{1}{2}}e^{-i\frac{2}{a}%
mx\sinh\frac{a\Delta\tau}{2}}, \label{G11}%
\end{equation}
and
\begin{equation}
G^{(12)}(\Delta\tau)=G^{(21)}(\Delta\tau)=\frac{m}{4L\pi^{2}}\int_{1}^{\infty
}dx\sin\left(  mL\left(  x^{2}-1\right)  ^{\frac{1}{2}}\right)  e^{-i\frac
{2}{a}mx\sinh\frac{a\Delta\tau}{2}}. \label{G12}%
\end{equation}

Their Fourier transformation,
\begin{align}
F^{(11)}(\omega)  &  =F^{(22)}(\omega)=\frac{\omega}{\pi}\frac{e^{\pi\omega
/a}}{\pi\omega/a}\int_{m/a}^{\infty}\sqrt{x^{2}-m^{2}/a^{2}}K_{i2\omega
/a}(2x)dx,\nonumber\\
F^{(12)}(\omega)  &  =F^{(21)}(\omega)=\frac{\omega}{\pi}\frac{e^{\pi\omega
/a}}{\pi\omega/a}\int_{m/a}^{\infty}\frac{sin(aL\sqrt{x^{2}-m^{2}/a^{2}})}%
{aL}K_{i2\omega/a}(2x)dx.
\end{align}

Note that $F^{(11)}(\pm\omega)$ for both massless and massive field is
proportional to the excitation/deexcitation rate of the single detector, and
it satisfies the KMS condition \cite{kubo1986brownian,martin1959theory}
\begin{equation}
\frac{F^{(11)}(\omega)}{F^{(11)}(-\omega)}=e^{-\frac{\omega}{T}}%
=e^{-\frac{2\pi\omega}{a}}\label{KMS}%
\end{equation}
where $T=a/2\pi$ is the Unruh temperature perceived by the detector.

\subsection{Entanglement dynamics}

We study the entanglement dynamics of two uniformly accelerated two-level
detectors, which are weakly coupled to fluctuating massless and massive scalar
fields in the vacuum. For convenience, we work in the coupled basis%
\[
\{|G\rangle=|00\rangle,|A\rangle=\frac{1}{\sqrt{2}}(|10\rangle-|01\rangle
),|S\rangle=\frac{1}{\sqrt{2}}(|10\rangle+|01\rangle),|E\rangle=|11\rangle\},
\]
and obtain the following time evolution equations of the density matrix
elements \cite{ficek2002entangled}%
\begin{align}
\rho_{GG}^{\prime}  &  =-4\left(  A_{1}-B_{1}\right)  \rho_{GG}+2\left(
A_{1}+B_{1}-A_{2}-B_{2}\right)  \rho_{AA}+2\left(  A_{1}+B_{1}+A_{2}%
+B_{2}\right)  \rho_{SS},\nonumber\\
\rho_{EE}^{\prime}  &  =-4\left(  A_{1}+B_{1}\right)  \rho_{EE}+2\left(
A_{1}-B_{1}-A_{2}+B_{2}\right)  \rho_{AA}+2\left(  A_{1}-B_{1}+A_{2}%
-B_{2}\right)  \rho_{SS},\nonumber\\
\rho_{AA}^{\prime}  &  =-4\left(  A_{1}-A_{2}\right)  \rho_{AA}+2\left(
A_{1}-B_{1}-A_{2}+B_{2}\right)  \rho_{GG}+2\left(  A_{1}+B_{1}-A_{2}%
-B_{2}\right)  \rho_{EE},\nonumber\\
\rho_{SS}^{\prime}  &  =-4\left(  A_{1}+A_{2}\right)  \rho_{SS}+2\left(
A_{1}-B_{1}+A_{2}-B_{2}\right)  \rho_{GG}+2\left(  A_{1}+B_{1}+A_{2}%
+B_{2}\right)  \rho_{EE},\nonumber\\
\rho_{AS}^{\prime}  &  =-4\left(  A_{1}+iD\right)  \rho_{AS},\quad\rho
_{SA}^{\prime}=-4\left(  A_{1}-iD\right)  \rho_{SA},\nonumber\\
\rho_{GE}^{\prime}  &  =-4A_{1}\rho_{GE},\quad\rho_{EG}^{\prime}=-4A_{1}%
\rho_{EG}, \label{DRHO}%
\end{align}
where $\rho_{IJ}=\langle I|\rho|J\rangle,I,J\in\{G,E,A,S\}$, $\rho
_{IJ}^{\prime}=\frac{\partial\rho_{IJ}}{\partial\tau}$, and $A_{1}\equiv
A^{(11)}=A^{(22)}, A_{2}\equiv A^{(12)}=A^{(21)}, B_{1}\equiv B^{(11)}%
=B^{(22)}, B_{2}\equiv B^{(12)}=B^{(21)}$.

Moreover, we just consider the effect of environment (the vacuum field) on
quantum entanglement between two accelerated detectors, so the Hamiltonian for
any single detector and vacuum contribution terms can be neglected, and we
just need to consider the effect of dissipator $\mathcal{D}[\rho(\tau)]$. In
order to eliminate the interaction between the two detectors caused by the
environment, we let $D=0$ in Eq. (\ref{DRHO}). The computation of the additive
environment leading term can be found in Ref. \cite{chen2022entanglement}.

\subsection{Entanglement generation}

\begin{figure}[ptb]
\centering
\includegraphics[width=0.6\linewidth]{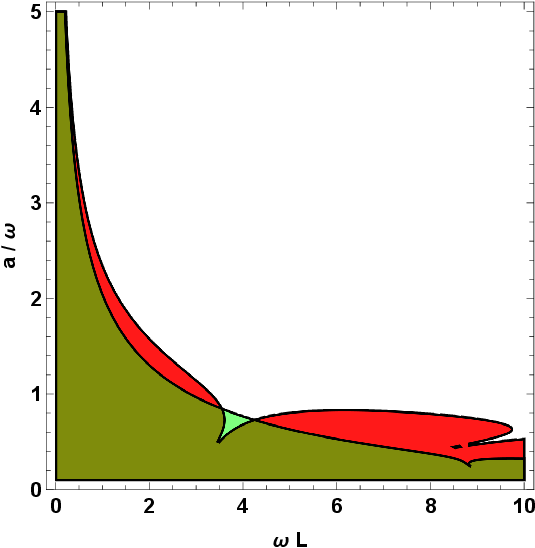} \caption[entanglement]{Parameter
region $(\omega L, a/\omega)$ within which entanglement generation is possible
for a uniformly accelerated two-atom system initially prepared in the state
$|g\rangle$. The red (green) region is related to the massless (massive)
field, and the brown-green region is the overlapping area. For the massive case,
the parameter is taken as $m/\omega=1$.}%
\label{Fig1}%
\end{figure}

We investigate the entanglement generation between two initially separable
uniformly accelerated detectors, and focus on the environment-induced
interaction. According to the evolution equation of the density matrix
elements (\ref{DRHO}), the density matrix elements affected by the
environment-induced interatomic interaction in the coupled basis are
$\rho_{AS}$ and $\rho_{SA}$.

The concurrence which quantifies the amount of quantum entanglement of the
bipartite entanglement state in the X-form state can be expressed as
\cite{wootters1998entanglement,tanas2004entangling}
\begin{equation}
C[\rho(\tau)]=max\{0,K_{1}(\tau),K_{2}(\tau)\},
\end{equation}
where
\begin{align}
K_{1}(\tau)=  &  \sqrt{[\rho_{AA}(\tau)-\rho_{SS}]^{2}-[\rho_{AS}-\rho
_{SA}]^{2}}\nonumber\\
&  -2\sqrt{|\rho_{GG}(\tau)\rho_{EE}(\tau)|},\nonumber\\
K_{2}(\tau)=  &  -\sqrt{[\rho_{AA}(\tau)+\rho_{SS}]^{2}-[\rho_{AS}+\rho
_{SA}]^{2}}\nonumber\\
&  +2|\rho_{GE}(\tau)|.
\end{align}

To calculate the entanglement dynamics of the system, assume that the initial
state of the two-atom system is $|10\rangle$ ($\rho_{AS}(0)=\rho_{SA}%
(0)=\frac{1}{2}$). In the environment-induced interaction, the element
$\rho_{GE}$ of the density matrix will remain at zero. So the value of
concurrence becomes $C[\rho(\tau)]=max\{0,K_{1}(\tau)\}$. Entanglement can be
generated at the neighborhood of the initial time $\tau=0$ when $K_{1}%
^{\prime}(0)>0$. The expression for $K_{1}^{\prime}(0)$ is given by
\begin{equation}
K_{1}^{\prime}(0)=4|A_{2}|-4\sqrt{A_{1}^{2}-B_{1}^{2}}.
\end{equation}

Fig. 1 shows the parameter space for entanglement generation when the
environment-induced interaction between the detectors accelerated in the
vacuum is considered. The estimated area of the entanglement region for
massless and massive fields are $12.6$ and $10.1$, respectively. It means that
the parameter space for entanglement generation shrinks when the mass of the
field is taken into account.

\subsection{Entanglement degradation}

Now, we discuss the degradation process of entanglement. The entanglement
between the two detectors will gradually decrease after generation. From the
previous sections, we know that three main factors affect the generation of
entanglement, which is acceleration, field mass, and the spatial distance
between detectors. In this paper, we consider the influence of acceleration
and field mass on the detectors' ability to extract entanglement when the
spatial distance is fixed.

\begin{figure}[ptb]
\centering
\includegraphics[width=0.45\linewidth]{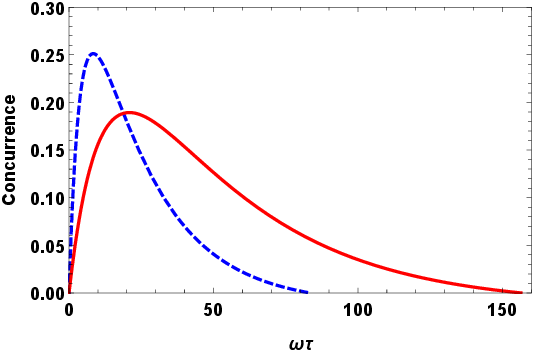}\qquad
\includegraphics[width=0.45\linewidth]{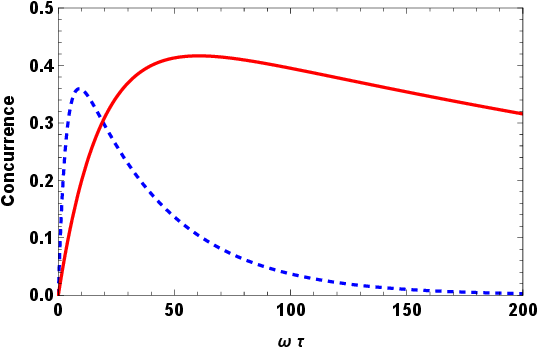} \caption[entanglement]%
{Concurrence as a function of interaction time $\tau$. The blue dotted line
shows the variation of entanglement generation in the dynamical evolution
of the detectors in a massless scalar field. The red solid line represents the
case in the massive field. Moreover, the left figure corresponds to $a/\omega=1$, and
the right figure corresponds to $a/\omega=0.2$. The distance between two detectors is
taken as $\omega L=1$.}%
\label{Fig2}%
\end{figure}

Fig. 2 shows the evolution of the entanglement harvesting of the detectors in
vacuum. Regardless of the influence of the field mass and acceleration, the
entanglement decreases with the increase of time after a sufficient
interaction between detectors and vacuum. The amount of entanglement will
first increase and then decrease with time. The entanglement drops to zero
when the interaction time is long enough. Firstly, focus on the left panel of
Fig. 2, where the red dotted line represents the entanglement evolution with
the paramenter $m/\omega=1$. When the mass $m=0$, the change of entanglement
returns to the case for the massless field. In particular, the presence of
field mass can slow down entanglement degradation, which may be related to the
time delay effect caused by field mass \cite{Zhou2021nyv}. The time delay
effect means, compared with the massless field, the entanglement evolution in
the massive case is generally slower. The right panel of Fig. 2 explains the
entanglement evolution process when the acceleration is small. Compared with
the left figure, the entanglement generation in the dynamical evolution of the
detectors increases for both massless and massive fields.

\section{Circular motion}

In this section, we will analyze the entanglement dynamics for two atoms in
circular motion, and compare them with the uniform acceleration case.

\begin{figure}[ptb]
\centering
\includegraphics[width=0.6\linewidth]{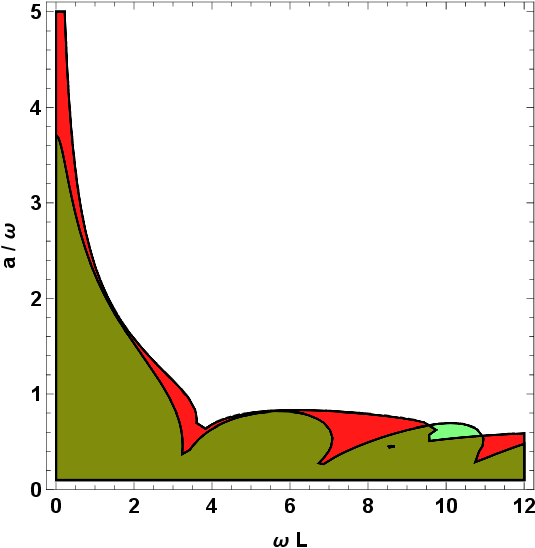} \caption{The parameter regions ($\omega L, a/\omega$) for
entanglement generation in the dynamical evolution of the detectors along the
trajectories of linearly accelerated motion and circular motion. The green
region corresponds to circular motion and the red region corresponds to
uniformly accelerated linear motion. The brown region is the part where these
two cases overlap. The parameters are taken as $\omega R=1, m/\omega=0.01$.}%
\label{Fig3}%
\end{figure}

We first give the trajectory of circular motion,
\begin{align}
t_{1}  &  =\gamma\tau,x_{1}=R\cos(\omega_{D}\gamma\tau),y_{1}=R\sin(\omega
_{D}\gamma\tau),z_{1}=0,\nonumber\\
t_{2}  &  =\gamma\tau,x_{2}=R\cos(\omega_{D}\gamma\tau),y_{2}=R\sin(\omega
_{D}\gamma\tau),z_{2}=L, \label{trajectory_circular}%
\end{align}
where $R$ is the radius of the circular trajectory in a plane parallel to the
xy-plane, $\omega_{D}$ is the angular velocity which can be either positive or
negative in the circular motion, and $\gamma=1/\sqrt{1-R^{2}\omega_{D}^{2}}$
denotes the Lorentz factor. In the detector's frame, the magnitude of
acceleration satisfied $a=\gamma^{2}\omega_{D}^{2}R=\gamma^{2}v^{2}/R$ with
the magnitude of linear velocity $v=|\omega_{D}|R<1$. Substituting the
trajectory into the Wightman function, we obtain%
\begin{align}
G^{11}(\Delta\tau)  &  =-\frac{1}{4\pi^{2}}\frac{1}{(\gamma\Delta
\tau-i\epsilon)^{2}-4R^{2}\sin^{2}(\gamma\omega_{D}\Delta\tau/2)},\nonumber\\
G^{12}(\Delta\tau)  &  =-\frac{1}{4\pi^{2}}\frac{1}{(\gamma\Delta
\tau-i\epsilon)^{2}-4R^{2}\sin^{2}(\gamma\omega_{D}\Delta\tau/2)-L^{2}}.
\label{Wightman_circular}%
\end{align}

Then, substituting the Wightman function into Eq. (\ref{GF}), we get
\begin{align}
F^{(11)}(\omega)  &  =F^{(22)}(\omega)=-\frac{1}{4\pi^{2}}\int_{-\infty
}^{\infty}d\Delta\tau e^{i\omega\Delta\tau}\frac{1}{(\gamma\Delta
\tau-i\epsilon)^{2}-4R^{2}\sin^{2}(\gamma\omega_{D}\Delta\tau/2)},\nonumber\\
F^{(12)}(\omega)  &  =F^{(21)}(\omega)=-\frac{1}{4\pi^{2}}\int_{-\infty
}^{\infty}d\Delta\tau e^{i\omega\Delta\tau}\frac{1}{(\gamma\Delta
\tau-i\epsilon)^{2}-4R^{2}\sin^{2}(\gamma\omega_{D}\Delta\tau/2)-L^{2}}.
\label{Fourier_circular}%
\end{align}

For the massive field case, the Wightman function is
\cite{Maeso2022Entanglement}
\begin{equation}
G(x,x^{\prime})=\frac{m}{4\pi^{2}}\frac{1}{[-(\Delta t-i\epsilon)^{2}%
+|\Delta\mathbf{x}|^{2}]^{1/2}}K_{1}(m\sqrt{-(\Delta t-i\epsilon)^{2}%
+|\Delta\mathbf{x}|^{2}}).
\end{equation}
We can also use the circular motion trajectory, and get the Wightman function
for the circular motion case
\begin{align}
G^{11}(\Delta\tau) &  =\frac{m}{4\pi^{2}}\frac{1}{[-(\gamma\Delta
\tau-i\epsilon)^{2}+4R^{2}\sin^{2}(\gamma\omega_{D}\Delta\tau/2)]^{1/2}%
}\nonumber\\
&  \times K_{1}(m\sqrt{-(\gamma\Delta\tau-i\epsilon)^{2}+4R^{2}\sin^{2}%
(\gamma\omega_{D}\Delta\tau/2)}),\nonumber\\
G^{12}(\Delta\tau) &  =\frac{m}{4\pi^{2}}\frac{1}{[-(\gamma\Delta
\tau-i\epsilon)^{2}+4R^{2}\sin^{2}(\gamma\omega_{D}\Delta\tau/2)+L^{2}]^{1/2}%
}\nonumber\\
&  \times K_{1}(m\sqrt{-(\gamma\Delta\tau-i\epsilon)^{2}+4R^{2}\sin^{2}%
(\gamma\omega_{D}\Delta\tau/2)+L^{2}}).
\end{align}
Their Fourier transformations
\begin{align}
F^{11}(\omega) &  =\frac{m}{4\pi^{2}}\int_{-\infty}^{\infty}d\Delta\tau
e^{i\omega\Delta\tau}\frac{1}{[-(\gamma\Delta\tau-i\epsilon)^{2}+4R^{2}%
\sin^{2}(\gamma\omega_{D}\Delta\tau/2)]^{1/2}}\nonumber\\
&  \times K_{1}(m\sqrt{-(\gamma\Delta\tau-i\epsilon)^{2}+4R^{2}\sin^{2}%
(\gamma\omega_{D}\Delta\tau/2)}).\nonumber\\
F^{12}(\omega) &  =\frac{m}{4\pi^{2}}\int_{-\infty}^{\infty}d\Delta\tau
e^{i\omega\Delta\tau}\frac{1}{[-(\gamma\Delta\tau-i\epsilon)^{2}+4R^{2}%
\sin^{2}(\gamma\omega_{D}\Delta\tau/2)+L^{2}]^{1/2}}\nonumber\\
&  \times K_{1}(m\sqrt{-(\gamma\Delta\tau-i\epsilon)^{2}+4R^{2}\sin^{2}%
(\gamma\omega_{D}\Delta\tau/2)+L^{2}}).\label{Maasive Fourier_circular}%
\end{align}
The expressions in Eqs. (\ref{Fourier_circular}) and
(\ref{Maasive Fourier_circular}) do not have analytic form, so we need to
perform them numerically.

The entanglement region for the circular motion case is shown in Fig. 3.
Comparing it with Fig. 1, we find that the shapes of the entanglement region
for circular and uniform line acceleration are similar, but the estimated area
of the entanglement region for the circular motion is smaller
than the uniform line acceleration case.

\begin{figure}[ptb]
\centering
\includegraphics[width=0.45\linewidth]{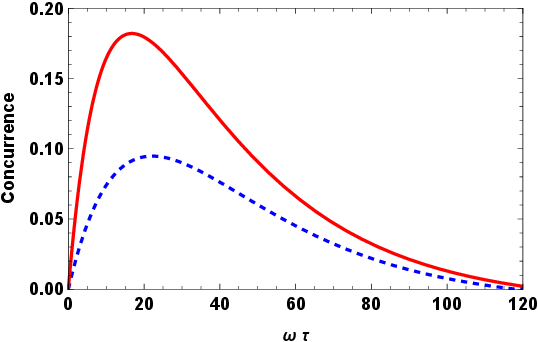}\qquad
\includegraphics[width=0.45\linewidth]{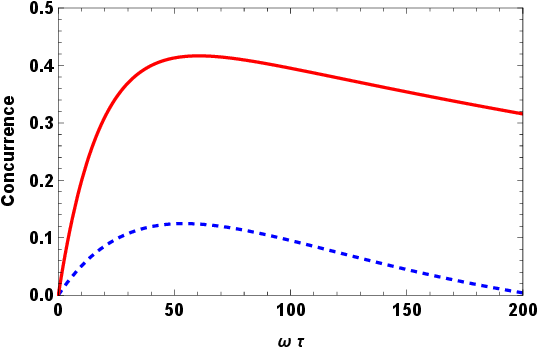} \caption{Concurrence as a
function of interaction time $\tau$. Solid and dashed lines stand for uniform
and circular cases respectively. The left figure and right figure stand for
$a/\omega=1.2$ and $a/\omega=0.2$ respectively.
Other parameters are $m/\omega=1$, $\omega R=1$, and $\omega L=1$.}%
\label{Fig4}%
\end{figure}

Fig. 4 shows the entanglement dynamics for the circular motion case. For the
same energy gap $\omega$ and distance $L$, the amount of entanglement
generated from the circular motion is less than the entanglement generated
from the uniform acceleration motion. When the acceleration $a$ is smaller, this
difference becomes bigger, because both cases approach the entanglement
generation for the static case.

\section{Anti-Unruh phenomenon and field mass}

From the right panels of Fig. 2 and Fig. 4, the detectors in the massive field
can harvest more entanglement when the acceleration is small. As discussed in
some previous studies, the anti-Unruh can increase entanglement among
entangled states for both bipartite states \cite{li2018would} and multi-body
states \cite{pan2020influence}, which makes us wonder if the anti-Unruh effect
can help generating entanglement between detectors in their dynamical
evolution process \cite{liu2022does}. To solve this problem, we analyze the
conditions for the anti-Unruh effect in detail using the methods of Refs.
\cite{garay2016thermalization,wu2023conditions}.

\begin{figure}[ptb]
\centering
\includegraphics[width=0.7\linewidth]{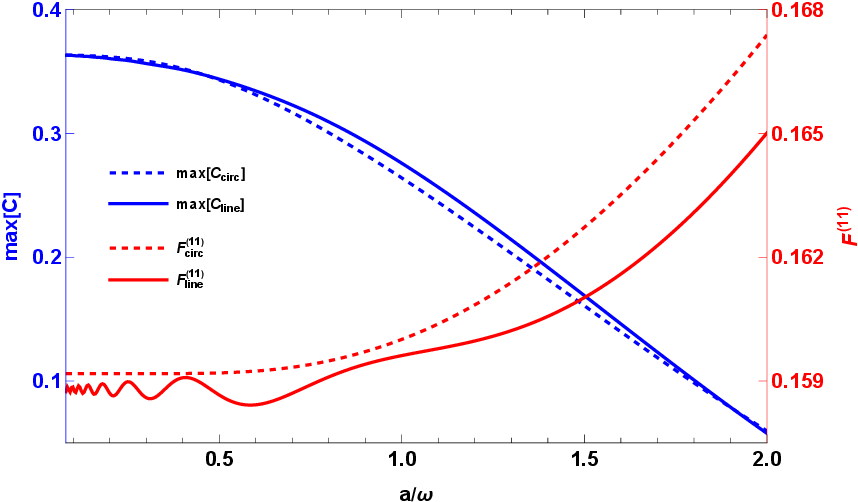} \caption{Maximum value of
concurrence and transition rate as a function of $a$. The red (blue)
solid line corresponds to the transition rate (the entanglement maximum) for
the case of the linear acceleration, and the red (blue) dashed line
corresponds to the transition rate (the entanglement maximum) for the case of
the circular motion. The parameters are taken as $m/\omega=0.1$, $\omega R=1$, and $\omega L=1$.}%
\label{Fig5}%
\end{figure}

Figure 5 illustrates the change in detector transition rate and entanglement
maxima (peaks at a given acceleration in Fig. 2 and 4) with acceleration for
linearly accelerated and circular motions. For linearly accelerated motion,
the fluctuations in the transition rate at small accelerations demonstrate the
presence of the weak anti-Unruh effect. However, this fluctuation has no
effect on the degradation of entanglement. For the circular motion, the weak
anti-Unruh effect is absent and has no effect on the entanglement degradation.
Therefore the Unruh and anti-Unruh effects are not necessarily related to the
degradation of entanglement.

Consider two different limits for the field mass: the large mass limit and the
small mass limit. Note that the small mass limit is not $m\rightarrow0$, and
the large one is not $m\rightarrow\infty$. A detailed discussion of these two
limits can be found in Ref. \cite{takagi1986vacuum,garay2016thermalization}.

Let us begin with the large mass limit. In order to show the dependence of the
field mass more clearly, the Fourier transformation for the green function of
the massive scalar field is first calculated. Substitute Eq. (\ref{G11}) into
Eq. (\ref{GF}) to get,%
\begin{equation}
F^{(11)}=\frac{m^{2}e^{\frac{\omega\pi}{a}}}{a\pi^{2}}\int_{1}^{\infty}%
\sqrt{x^{2}-1}K_{\frac{i2\omega}{a}}[\frac{2mx}{a}]dx.\label{FG}%
\end{equation}
The asymptotic expansion of the Bessel function for large values of its
argument is given as\textit{ }%
\begin{equation}
K_{\nu}(z)\sim\left(  \frac{\pi}{2z}\right)  ^{1/2}e^{-z}\left\{  1+\frac
{4\nu^{2}-1}{8z}+\cdots\right\}  ,\label{bessel}%
\end{equation}
and we use only the leading order term in our calculation, that is the first
term in Eq. (\ref{bessel}). It can be justified under the condition
\begin{equation}
\frac{(\omega/a)^{2}+1}{m/a}\ll1.\label{LC}%
\end{equation}

Thus, we get the following response function in the large mass limit,%
\begin{equation}
F_{lm}^{(11)}\approx\frac{m^{3/2}e^{\frac{\omega\pi}{a}}}{2\sqrt{a}\pi^{3/2}%
}\int_{1}^{\infty}\frac{\sqrt{x^{2}-1}}{\sqrt{x}}e^{-2mx/a}dx.\label{FGL}%
\end{equation}

We are at this point going to expand the second class of modified Bessel
functions to the second order of the curly brackets in Eq. (\ref{bessel}), and
the asymptotic behavior of the response function in the limit of a small mass
is given by%
\begin{equation}
F_{sm}^{(11)}\approx\frac{m^{3/2}e^{\frac{\omega\pi}{a}}}{2\sqrt{a}\pi^{3/2}%
}\int_{1}^{\infty}\frac{\sqrt{x^{2}-1}}{\sqrt{x}}e^{-2mx/a}(1-\frac
{16\omega^{2}+a}{16mx})dx.\label{FGS}%
\end{equation}
This is justified under the condition
\begin{equation}
\frac{(m/a)^{2}}{\left[  1+(\omega/a)^{2}\right]  ^{1/2}}\ll1,\label{SC}%
\end{equation}
which is roughly, but not quite, the opposite of Eq. (\ref{LC}).

\begin{figure}[ptb]
\centering
\includegraphics[width=0.6\linewidth]{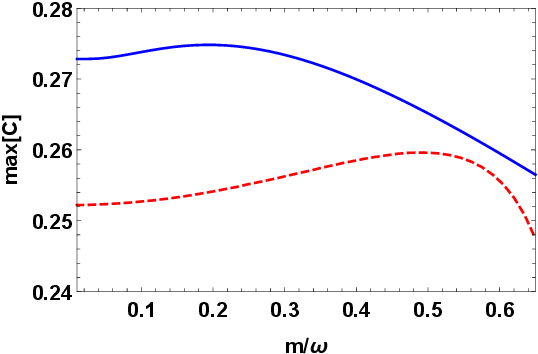} \caption{Maximum value of
concurrence as a function of field mass. The blue solid line corresponds to
linear accelerated motion and the red dashed line corresponds to circular
motion. The parameters are taken as $a/\omega=1$, $\omega R=1$, and $\omega L=1$.}%
\label{Fig6}%
\end{figure}

Figure 6 illustrates the change of the entanglement maximum with the field
mass for linear acceleration and circular motions. The entanglement captured
by the detector increases at small accelerations, but decreases as the mass
increases further. In order to check whether the anti-Unruh effect plays a
role in this process, we calculated the KMS temperatures perceived by the
detector in uniform acceleration and the detector in circular motion.

The KMS temperatures of the field can be calculated from the ratio between the
excitation and deexcitation rates of the detectors \cite{takagi1986vacuum},
\begin{equation}
T=-\frac{\omega}{\log\left[  \mathcal{R}(\omega)\right]  }, \label{edr}%
\end{equation}
where $R(\omega)=\frac{F^{(11)}(\omega)}{F^{(11)}(-\omega)}$. The response
function is also proportional to the transition rate of the detector, when it
interacts with a (3+1) dimensional massive field. So we get,
\begin{equation}
T_{lm}=T_{sm}=\frac{a}{2\pi}. \label{EDR temperature}%
\end{equation}
Thus, we can see that the strong anti-Unruh effect does not occur in both the
small mass limit and the large mass limit.

\begin{figure}[ptb]
\centering
\includegraphics[width=0.45\linewidth]{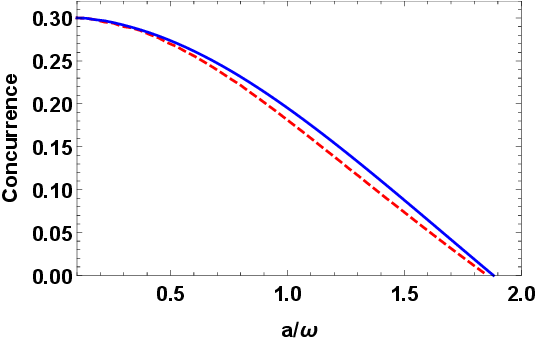}\qquad
\includegraphics[width=0.45\linewidth]{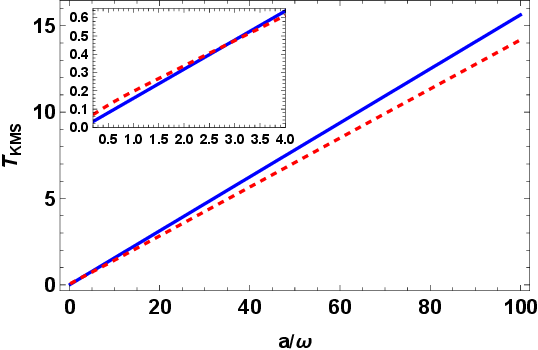} \caption{The left figure: The
entanglement for a certain time interval as a function of
acceleration. The parameters are taken as $\tau\omega=20$, $\omega R=1$, and $\omega L=1$. The
right figure: KMS temperature as a function of acceleration. The parameters
are taken as $\omega R=1$ and $m/\omega =1$. The blue curve corresponds to the linear
acceleration and the red dashed line corresponds to the circular motion.}%
\label{Fig7}%
\end{figure}

The left-hand panel of Fig. 7 shows the change of entanglement with
acceleration for a fixed spatial interval and evolution time. The detector
extracts more entanglement for two atoms in the uniform acceleration than that
in the circular motion. Based on the difference of the temperature induced by
acceleration for circular and linear acceleration motions
\cite{Biermann2020Unruh} and the effect of thermal field on the entanglement
of quantum entangled states \cite{Dai2015ota,barman2021radiative}. We
understand that circular motion produces less entanglement because circular
motion produces a higher Unruh temperature at smaller accelerations.

The right-hand panel of Fig. 7 illustrates the relation between the KMS
temperature and the acceleration for both circular and linear motion cases.
Remarkably, the KMS temperature for linear accelerated detectors is
proportional to the acceleration $T=a/2\pi$ and is irrelevant with the field
mass. There is no analytic expression for the KMS temperature for circular
accelerated detectors, but it still increases with the increasing acceleration
\cite{Biermann2020Unruh}. So there is no appearance of the strong anti-Unruh
effect in the whole process.

On the other hand, when the acceleration is small, the circularly accelerated
detectors feel a higher temperature compared with the linear acceleration
case; when the acceleration is large, detectors with linearly accelerated
motions feel higher temperature. The results are consistent with previous
results like that in Ref. \cite{Biermann2020Unruh}. Therefore, during
entanglement harvesting, the weak anti-Unruh effect has an impact on the
entanglement dynamics of the detector, but the strong anti-Unruh effect does
not occur and does not contribute.

\section{Conclusion}

We explore the entanglement dynamics between two detectors undergoing uniform
linear acceleration and circular motion within a massive scalar field. First,
we find that under linear acceleration, detectors are more readily entangled
in a massless scalar field compared to a massive scalar field. Additionally,
we analyze the evolution of entanglement between detectors in conjunction with
their intrinsic time scales. Our findings reveal that detectors exhibit
greater entanglement capture at lower accelerations, observed in both massive
and massless scalar field scenarios. Furthermore, we observe a noticeable
time-delay effect attributable to mass production.

Next, we analyze the entanglement generation process between detectors during
circular motion. Despite the similarity in the entanglement-generating
region's profile compared to linear acceleration, circular motion is less
conducive to generating entanglement compared to linear acceleration. We then
compare the entanglement dynamics between detectors experiencing acceleration
during circular motion versus linear acceleration. Our results indicate that
detectors in linear acceleration yield more entanglement when the acceleration
is small. This can be attributed to the linearly accelerating detector
perceiving a lower KMS temperature under mild acceleration. Similarly, we find
that in circular motion scenarios, detectors also harvest more entanglement
with lower acceleration, regardless of whether the scalar field is massive or massless.

Finally, we investigate the relationship between entanglement generation in
the dynamical evolution of the detectors and the mass of the field. Our
findings reveal a certain increase in the maximum entanglement between
detectors when the field mass is small. However, this increase is not
attributed to the strong anti-Unruh effect. We begin by defining the
anti-Unruh effect and determine that there is no presence of the strong
anti-Unruh effect throughout the process. However, the weak anti-Unruh effect
does impact the entanglement dynamics of the detector. Interestingly, the
influence of the weak anti-Unruh effect differs from the conventional
understanding (where entanglement increases), as the entanglement between
detectors does not exhibit a strictly monotonic dependence on acceleration.

\acknowledgments

Pan and Yan contributed equally to this work. This work is supported by
National Natural Science Foundation of China (NSFC) with Grant No. 12375057
and the Fundamental Research Funds for the Central Universities, China
University of Geosciences (Wuhan).

\end{document}